\documentclass[12pt]{article}
\usepackage{graphicx,amsmath}
\usepackage{graphicx,amsmath,lineno,color}

\usepackage{units}
\usepackage{caption} 
\usepackage[backend=biber,sorting=none,natbib=true,doi=true]{biblatex} 
\renewbibmacro{in:}{}
\usepackage[hidelinks,colorlinks=true,linkcolor=blue,urlcolor=blue,citecolor=blue]{hyperref}
\hypersetup{colorlinks=true}

\newlength{\dinwidth}
\newlength{\dinmargin}
\def\lapproxeq{\lower .7ex\hbox{$\;\stackrel{\textstyle                                                    
<}{\sim}\;$}}                                                    
\def\gapproxeq{\lower .7ex\hbox{$\;\stackrel{\textstyle                                                    
>}{\sim}\;$}}                                                    
\def\be{\begin{equation}}                                                    
\def\ee{\end{equation}}                                                    
\def\bea{\begin{eqnarray}}                                                    
\def\eea{\end{eqnarray}}

\def\sh{\hat s}
\def\sh2{{\hat s}^2}

\begin{document}                                                    
\titlepage                                                    
\begin{flushright}                                                    
\today \\                                                    
\end{flushright} 
\vspace*{0.5cm}

\begin{center}                                                    
{\Large \bf Comments on low mass dissociation at the LHC in the context of the discrepancy between the ATLAS and TOTEM measurements of $\sigma_{tot}$}\\

\vspace*{1cm}

P.~Grafstr\"om  \\                                                   
                                                   
\vspace*{0.5cm}                                                    
 Universit\`a di Bologna, Dipartimento di Fisica , 40126 Bologna, Italy\\
                        
\vspace*{1cm}                                                    
 
\begin{abstract}
The cross section for low mass dissociation at LHC energies is estimated in a partly data driven approach. The result is compared to the Monte Carlo estimate  from the QGSJET-II-03 model used by the TOTEM experiment in the determination of $\sigma_{tot}$ via the luminosity-independent method. Significant differences are found and possible consequences are explored and discussed.  \end{abstract}

\end{center}

\vspace{1cm}

\section{Introduction}
\begin{figure}[h]
 \begin{center}
\includegraphics[width=0.6\textwidth]{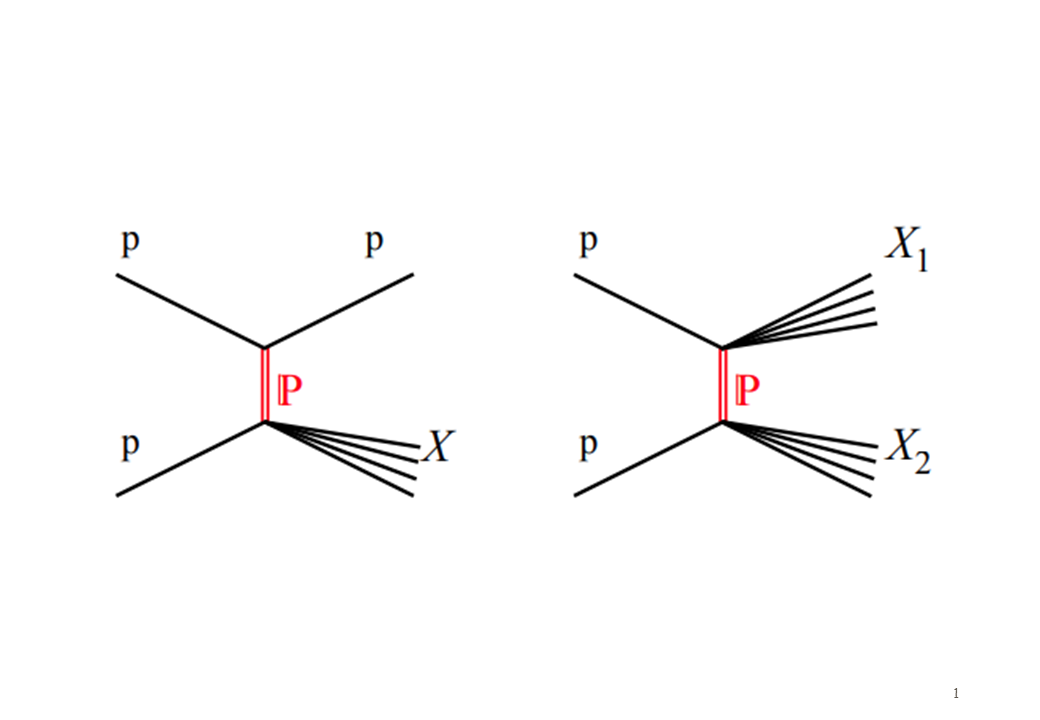}
\caption{\sf Pomeron exchange diagrams corresponding to single and double dissociation. The figure is adapted from Ref.~\cite{ALICE:2012fjm}.}
\label{fp1}
\end{center}
\end{figure}

With the term low mass diffraction or dissociation into a low mass system  we mean a low $p_{t}$ process of the type illustrated  in Fig.~\ref{fp1},  where the topologies single    and double dissociation   are shown.  These  topologies are characterized by the exchange of  the Pomeron in the $t$-channel or in the language of QCD  an exchange of a color singlet state of  gluons.  With low mass  in this context we refer  to processes where the mass of the dissociated system is less than $\approx $ 3--4 GeV for single dissociation and in the case of double dissociation both systems have a mass less than $\approx $ 3--4 GeV.

 The cross section for these type of reactions are both hard to measure and  challenging to theoretically estimate. In general it is difficult to measure diffraction at high energy colliders but especially for diffractive systems with a low mass. For low mass systems a large fraction of the diffractively produced particles are emitted in the very forward direction and lost in the beam pipe. Also the  theoretical estimates of low mass diffraction   are  notorious difficult and inexact.  Often the Good-Walker approach \cite{Good:1960ba} is used  but such an approach is associated with  a number of uncertainties.

The size of the low mass cross section has of course an interest in itself but there is also an interest related to the measurements of the total cross section ($\sigma_{tot}$).   Actually, the uncertainties related to low mass dissociation constitute   the largest uncertainty of the total cross section measurements by the TOTEM experiment using the so called luminosity-independent method \cite{TOTEM:2013vij}\cite{TOTEM:2012oyl}\cite{TOTEM:2017asr}.

Using the luminosity-independent method the total cross section is given by:

\begin{equation}
\label{opt2}
 \sigma_{tot}=~  \frac{16\pi\frac{}{}}{1+\rho^{2}}^{}_{ }\frac{(dN_{el}/dt)_{t=0} }{(N_{_{}el}+N_{inel})}_{}
\end{equation}

where $N_{el}$ and $N_{inel}$  stand for the elastic and inelastic rate,$  ~(dN_{el}/dt)_{t=0}$  is the elastic differential rate extrapolated to $t=0$  and $\rho$ is the ratio of the real  to  imaginary part of the elastic scattering amplitude in the forward direction. $N_{inel}$  is measured except  in the very forward direction and thus an estimate of the  rate  from  low mass dissociation is required. 

TOTEM uses Monte Carlo  program QGSJET-II-03 \cite{Ostapchenko:2004ss} to estimate how much is undetected between $ M_{x}\approx$ 3-4~  GeV and the proton mass. The mass of the diffractive system, $ M_{x}$, is related to the commonly used variable $\xi $ through  $\xi=M_{x}^{2}/s $ with s being the center-of-mass energy squared.

 The Monte Carlo estimates used by TOTEM are given in Table~\ref{tab1}. The authors of Refs.~\cite{TOTEM:2013vij}\cite{TOTEM:2012oyl}\cite{TOTEM:2017asr}
are aware of the difficulty to make a precise estimate and give  uncertainties representing half of the  correction.

\begin{table}[h!]
\centering
\caption{Corrections used by TOTEM for low mass dissociation}
\begin{tabular}{|c|c|c|c|c|}
\hline
$\sqrt{s}$  & Low mass limit & Correction & $\sigma_{inel}$ & Correction \\
TeV & GeV & \% of $\sigma_{inel}$ & mb & mb \\
\hline
7&3.4&4&72.9&2.9\\
8&3.6&4.8$\pm$2.4&74.7&3.6$\pm$1.8\\
13&4.6&7.1$\pm$3.55&79.5&5.6$\pm$2.8\\
\hline
\end{tabular}
\label{tab1}
\end{table}

The strong energy dependence in going from 7 TeV to 13 TeV i.e. from 4\% to 7\% appears somewhat unnatural to us given that the cross sections in the LHC regime rather vary as logarithms of the center-of mass energy or as a power law but with a small power of order $\approx 0.1.$

We asked ourselves if there is another way of estimating low mass dissociation  not relying exclusively upon Monte Carlo estimates but rather using at least a  partly  data driven approach. 

\section{Measurements of inelastic cross sections at 7 TeV and 13 TeV}

Table~\ref{tab2} and Table~\ref{tab3} summarize measurements of the inelastic cross sections at 
$\sqrt{s}$=7~TeV and at $\sqrt{s}=13$~TeV. There
are the measurements of the total inelastic cross section ($\sigma_{inel}=\sigma_{tot}-\sigma_{el}$ ) from ATLAS-ALFA and from TOTEM. In addition there are fiducial cross section measurements, in a limited kinematical range, from ALICE, ATLAS and CMS.
In the case of 7 TeV the measurements are for masses $M_{x}$ above 16 GeV, while in the case of 13 TeV the measurements are for masses  above 13 GeV. As can be seen there is good agreement between the different measurements of fiducial cross section in between the experiments. As also indicated in the tables one obtains by subtraction the inelastic cross section for masses below 16 GeV, respectively below 13 GeV. 

\begin{table}[h!]
\centering
\caption{Measurements of the inelastic  cross sections at 7 TeV }
\begin{tabular}{|l|c|l|c|c|}
\hline
Experiment& $\sigma_{inel}$ [mb]&Experiment & $\sigma_{inel}^{M_{x}>16~\mathrm{GeV}}$ [mb]&$\sigma_{inel}^{M_{x}<16~\mathrm{GeV}}$ [mb]\\
\hline
ATLAS-ALFA~\cite{ATLAS:2014vxr}&71.3$\pm$0.9&ALICE~\cite{ALICE:2012fjm}&62.1$\pm$2.4& \\
TOTEM~\cite{TOTEM:2013vij}&72.9$\pm$1.5&ATLAS~\cite{ATLAS:2011zrx}&60.3$\pm$2.1&\\
&&CMS~\cite{CMS:2012gek}&60.2$\pm$2.6&\\
\hline
 Weighted mean&71.8$\pm$0.7&&60.8$\pm$1.4&11.0$\pm$1.6\\
\hline
\end{tabular}
\label{tab2}
\end{table}

\begin{table}[h!]
\centering
\caption{Measurements of the inelastic cross sections at 13 TeV }
\begin{tabular}{|l|c|l|c|c|}
\hline
Experiment & $\sigma_{inel}$ [mb] & Experiment & $\sigma_{inel}^{M_{x}>13~\mathrm{GeV}}$ [mb] & $\sigma_{inel}^{M_{x}<13~\mathrm{GeV}}$ [mb]\\
\hline
ATLAS-ALFA~\cite{ATLAS:2022mgx}&77.4$\pm$1.1&ALICE&& \\
TOTEM~\cite{TOTEM:2017asr}&79.5$\pm$1.8&ATLAS~\cite{ATLAS:2016ygv}&68.1$\pm$1.4&\\
&&CMS~\cite{CMS:2018mlc}&67.5$\pm$1.8&\\
\hline
 Weighted mean&78.0$\pm$0.9&&67.8$\pm$1.1&10.2$\pm$1.4\\
\hline
\end{tabular}
\label{tab3}
\end{table}

\section{Estimates of low mass dissociation}

\begin{figure}[h]
 \begin{center}
\includegraphics[width=0.6\textwidth]{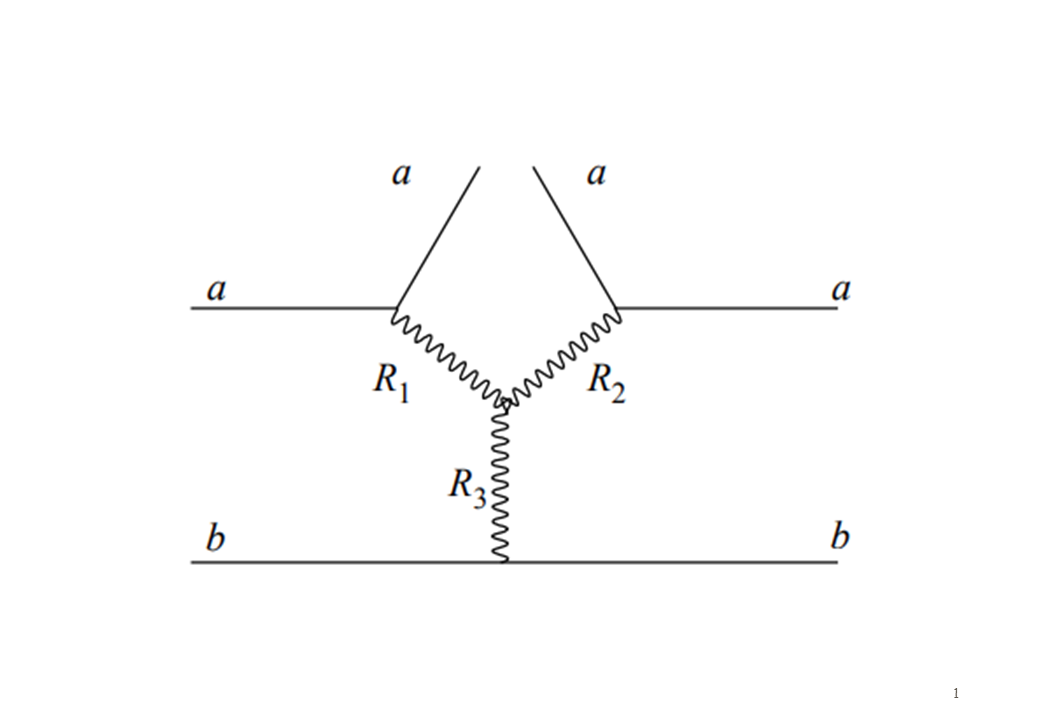}
\caption{\sf Triple-Reggeon Feynman diagram occurring in the calculation of the amplitude for single diffraction corresponding to the dissociation of hadron b in the interaction with hadron a. Each of the Reggeon legs can be a Pomeron or a secondary Reggeon. The figure is taken from Ref.~\cite{ALICE:2012fjm}.}
\label{fp2}
\end{center}
\end{figure}

 To use the data in Table~\ref{tab2} and Table~\ref{tab3}
 for an evaluation of  low mass dissociation one has to do some estimates of how much  of the cross section is to be found in the region between $M_{x}$ $\approx$ 13--16~ GeV and $M_{x}$ $\approx$ 3--4 GeV. Our main point in this note  is to argue that  such an estimate might be more reliable than an estimate of what is happening between the proton mass and $M_{x}$ $ \approx$ 3--4 GeV.
As mentioned in the introduction  there are many difficulties related to the  estimate of the cross section in the low mass region e.g. how many Good-Walker states are needed? On the other hand there is some kind of consensus that above the low mass region     
the so called triple Pomeron diagram, illustrated in Fig.~\ref{fp2},
dominates. At lower masses, on the contrary, it is not clear to what extent this diagram is important. At lower masses there are also uncertainties related to the importance of diagrams where one of the Pomerons  ($R_{3}$ in Fig.~\ref{fp2}) is replaced by a secondary Reggeon (see e.g. Ref.~\cite{Kaidalov:1973tc} or Ref.~\cite{Collins:1977jy}). 

At 13 TeV the situation is rather encouraging with respect to the cross section in the range 13 GeV $ >M_{x} > 4.1$ GeV. 
There is data from the CMS experiment measuring a cross section of 1.1 mb  on one side of the interaction point  corresponding to 2.2 mb for the two sides \cite{CMS:2018mlc}.
 There are also Monte Carlo calculations  for the same mass range given in Table~\ref{tab4} with values between 2 mb  and 3 mb \cite{CMS:2018mlc}.
Finally we have used the KMR model \cite{Khoze:2020plm} to estimate the cross section in this same mass range.
The result is 2.4 mb agreeing with the CMS measurements and the different Monte Carlo estimates.
Below we will use the 2.2 mb of the CMS measurement.

\begin{table}[h!]
\centering
\caption{Monte Carlo estimates of the fractional increase of $\sigma_{inel}$ in the region 13 GeV $ >M_{x} > 4.1$ GeV at $\sqrt{s}$=13 TeV. Taken from Ref.~\cite{CMS:2018mlc}.}
\begin{tabular}{|c|c|c|c|}
\hline
Monte Carlo &\% of  $\sigma_{inel}$& One side [mb]&Two sides [mb]\\
\hline
EPOS-LHC&1.76&1.2&2.4\\
QGSJETII-04&2.36&1.6&3.2\\
PYTHIA8 MBR&2.32&1.6&3.2\\
\hline

\end{tabular}
\label{tab4}
\end{table}

At 7 TeV there is no measurement in the region range 16 GeV $ >M_{x} > 3.4$ GeV but we can do several estimates. The result of different Monte Carlo estimates are shown in the appendix of  Ref.~\cite{CMS:2015inp} in a number of graphs. 
Looking at the graphs corresponding to the Monte Carlo programs listed in Table~\ref{tab4} we get values of the order of a couple of mb's. The KMR model \cite{Khoze:2020plm} gives 3 mb. There is also ATLAS data \cite{ATLAS:2012djz} measured just above $M_{x}$=16 GeV indicating 1~mb/unit of rapidity gap and using this below $M_{x}$=16 GeV would give 3.1 mb in 
the range 16~GeV~$ >M_{x} > 3.4$ GeV. In the following we will use 3 mb.

Putting together the information from the last column in Table~\ref{tab2} and Table~\ref{tab3} together 
with the estimates in this section we get by subtraction the cross section for low mass dissociation at 7 TeV and 13 TeV. 
The obtained values are given in Table~\ref{tab5} together with the values that has been used by TOTEM for comparison. 
Observe that the uncertainty on the estimated values are just the propagation of the uncertainty of the numbers used and 
does not include any uncertainties related to assumptions made to estimate the cross section in the
range $M_{x}\approx$3--4 GeV to $ M_{x}\approx$13--16 GeV. 
This means that the uncertainties certainly are underestimated. Also  observe that at 13 TeV we had to make  a small correction between the $M_{x}$=4.1 GeV from the CMS measurement as compared to the $M_{x}$=4.6 GeV used by TOTEM. The correction was estimated to 0.2 mb.

It must also be said that in the considerations above we have not made the distinction between  single dissociation and  double dissociation. The estimation given above from the KMR model contains only single dissociation. 
Double dissociation is thought to be of considerable less importance  being down with more than an order of magnitude 
relative to single dissociation~\cite{Khoze:2014aca}. This is also confirmed by the CMS data at 13 GeV discussed above which 
agree with the KMR estimate. 

\begin{figure}[h]
 \begin{center}
\includegraphics[width=\textwidth]{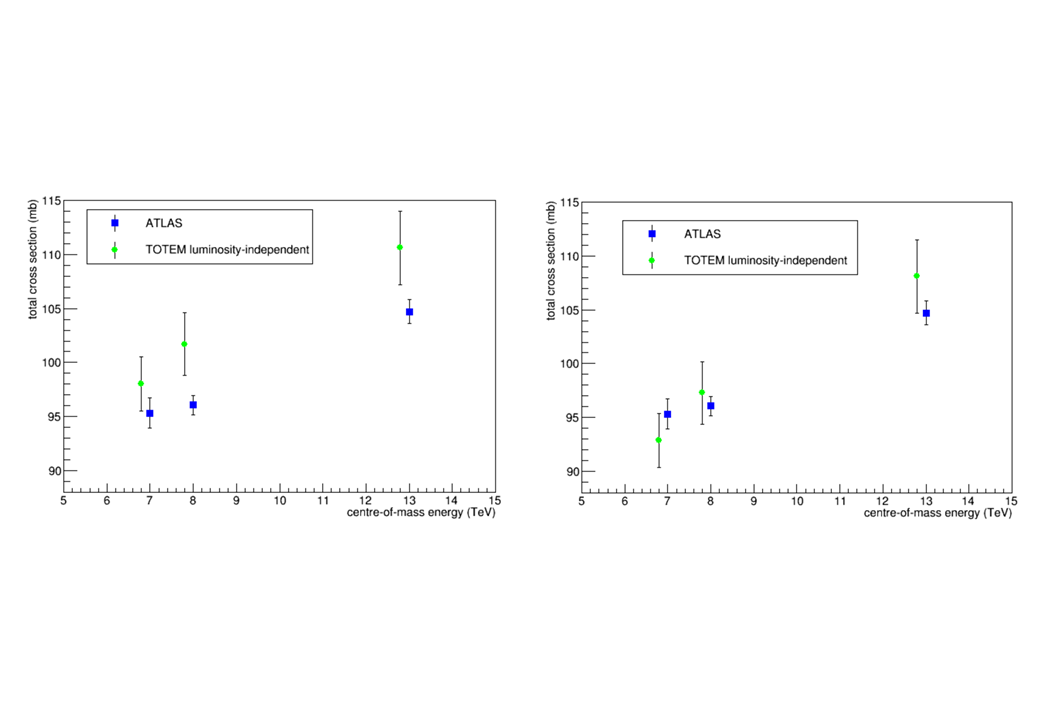}
\vspace{-3cm}
\caption{\sf Comparison of ATLAS and TOTEM measurements of $\sigma_{tot}$. To the left is shown the actual situation. 
To the right is shown what happens if the low mass estimates of this note are used.}
\label{fp3}
\end{center}
\end{figure}

\begin{figure}[h]
 \begin{center}
\includegraphics[width=0.7\textwidth]{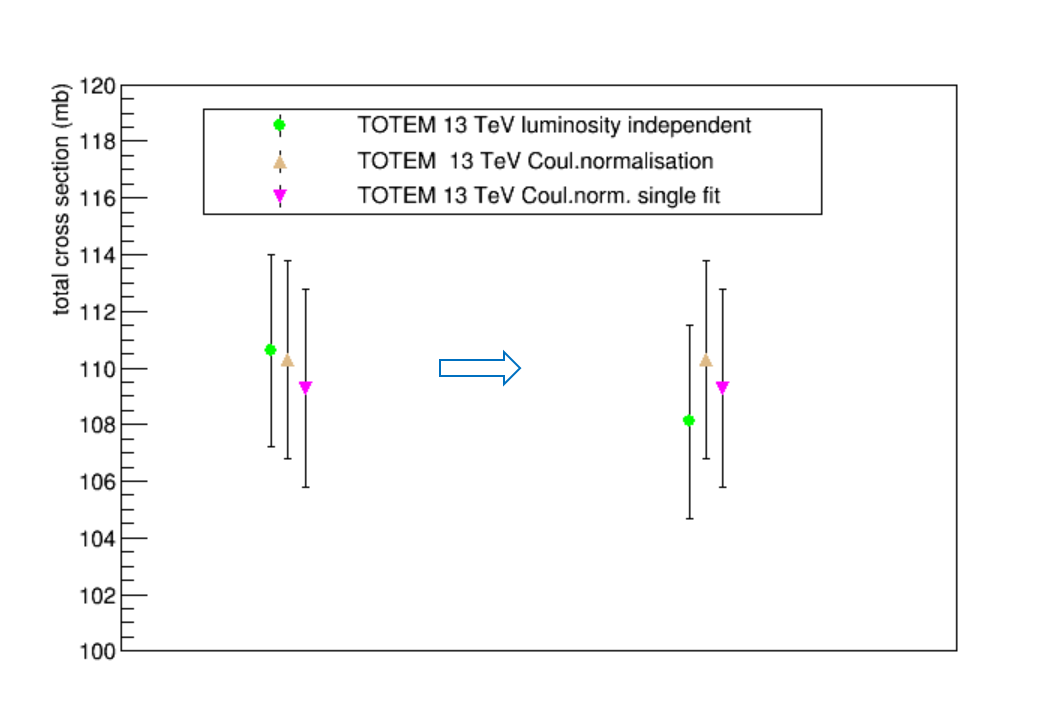}
\caption{\sf 
Comparison of TOTEM measurements of $\sigma_{tot}$ using Coulomb normalisation and using the luminosity-independent method. Two different approaches have been used to perform the Coulomb normalisation independently of the luminosity-independent method. Details are given in Ref.~\cite{TOTEM:2017sdy}. The figure shows, on the right side, the implications of the low mass corrections estimated in this note.} 
 \label{fp4}
\end{center}
\end{figure}
\begin{figure}[h]
 \begin{center}
\includegraphics[width=0.7\textwidth]{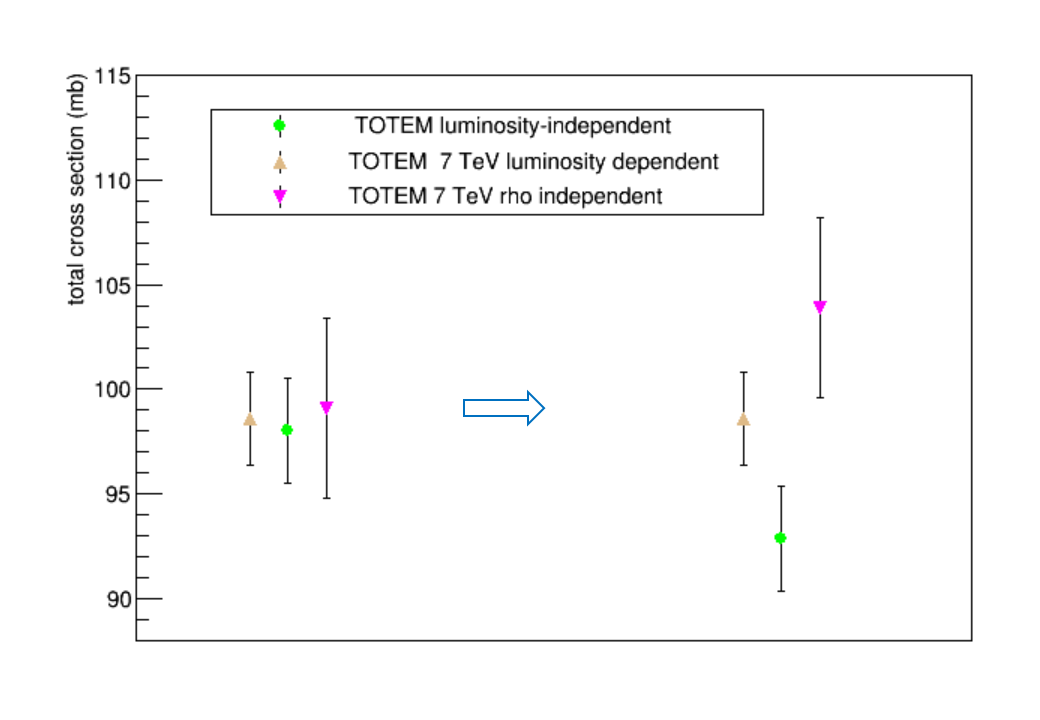}
\caption{\sf Comparison of TOTEM measurements of $\sigma_{tot}$ using three different methods before and after adjusting the low mass correction.} 
\label{fp5}
\end{center}
\end{figure}

In Table~\ref{tab5} it can be seen that at 7 TeV there is a striking difference between the estimate of low mass dissociation  used by TOTEM  and the present estimate. The difference is less pronounced at 13 TeV.
\begin{table}[h!]
\centering
\caption{Partly data driven  low mass dissociation cross section  compared to the Mont Carlo estimated used by TOTEM.}
\begin{tabular}{|c|c|c|c|}
\hline
$\sqrt{s}$ [TeV] & $ M_{x} < (GeV)$ & Low mass $\sigma_{inel}$ [mb]&Used by TOTEM [mb]\\
\hline
7&3.4&8.0$\pm$1.6&2.9\\
8&3.6&&3.6$\pm$1.8\\
13&4.6&8.2$\pm$1.4&5.6$\pm$2.8\\
\hline

\end{tabular}
\label{tab5}
\end{table}

\section{Consequences of using different   low mass corrections}
It is easy to adjust the TOTEM measurement for a given input in terms of low mass dissociation. This is what has been done in Fig.~\ref{fp3}.  This figure shows to the left the TOTEM  values of $\sigma_{tot}$  and to the right the adjusted TOTEM values calculated according to the low mass dissociation contribution estimated here. The original uncertainties given by TOTEM have been kept. There is no data available for 8 TeV and we have used the value obtained a 7 TeV.  The values of 
$\sigma_{tot}$  are compared to those of ATLAS~\cite{ATLAS:2014vxr,ATLAS:2022mgx,ATLAS:2016ikn} before and after the adjustment. One can observe that the original very significant difference between the two experiments transforms to a more or less normal fluctuation between two independent measurements.

It should be emphasised that this adjustment of the TOTEM values  is just an attempt to investigate  what the consequences are for different inputs of the low mass dissociation estimate and should by no means be considered as a recommended adjustment of the TOTEM data.

As will be seen in the next section there are also objections that can be raised against those adjusted values of  
$\sigma_{tot}$.   

\section{Discussion}
A priori, there are at least three objections that can be raised against the adjusted TOTEM values of $\sigma_{tot}$. 
We will discuss all three here below  in order below in order of increasing concern. 

\begin{enumerate}

\item  TOTEM  has used Coulomb normalisation at 13 TeV independently  of the luminosity-independent method~\cite{TOTEM:2017sdy}. The situation is summarised in Fig.~\ref{fp4}. The values of $\sigma_{tot}$
 of TOTEM using Coulomb normalisation are compared to the value from the luminosity-independent measurement before and after the low mass dissociation adjustment. As can be seen there is actually no issue here. The values  from Coulomb normalisation are compatible with both the  values of $\sigma_{tot}$
 obtained by TOTEM and the value of $\sigma_{tot}$
obtained here by adjusting the low mass contribution.
\item At an early stage of the 7 TeV analysis TOTEM  used a luminosity estimate   given to them by the  CMS experiment and having an uncertainty of 4\% \cite{TOTEM:2013lle}. TOTEM used this luminosity estimate to determine   $\sigma_{tot}$
with two different methods:  the luminosity-dependent method and the $\rho$-independent method \cite{TOTEM:2013vij}. In this case the situation 
 is more critical. The data points are  shown before and after the adjustment in Fig.~\ref{fp5}.  On one hand, using the original TOTEM values, the agreement is excellent. However after the adjustment the spread is significantly increased. Here the 4~\% luminosity measurement from CMS  plays a crucial role both for the luminosity-dependent method and the $\rho$-independent method. 
 A correction  of a couple of~\%  would not harm the compatibility before the adjustment and  make everything more compatible  after the adjustment.

\item At 7 TeV TOTEM has made a fiducial measurement of $\sigma_{inel}$ for masses $M_{x}$\textgreater3.4 GeV. 
From this measurement TOTEM has derived an upper limit of 6.31 mb at 95\% confidence level for the cross section for events with diffractive masses below 3.4 GeV at 7 TeV \cite{TOTEM:2013dtg}.
This is of course in   contrast  to the value found here and quoted in Table~\ref{tab5} of~ $8.0\pm1.6$ mb.  It is the same fiducial measurement that causes the rather high  value of $\sigma_{tot}$ after adjustment of the $\rho$-independent method  seen to the right in Fig \ref{fp5}.   Clearly there is a contradiction at some level between the estimates of low mass diffraction in this note and the TOTEM fiducial measurement at 7 TeV. On the other hand the contradiction between the TOTEM and ATLAS measurements of $\sigma_{tot}$ is even more striking and  we are exploring   possible explanations for  those discrepancies. 
\end{enumerate}

\section{Conclusion}
This note is intended only as an attempt  to evaluate the contribution of  low mass dissociation in a partly data driven approach and then investigate  possible consequences of such an estimation.

The overall conclusion of this note is:

 If it  turns out that low mass dissociation at LHC is somewhat underestimated then the discrepancy between the TOTEM and ATLAS measurement of $\sigma_{tot}$  would not be as dramatic as now but instead rather acceptable.\\

{\bf\Large Acknowledgements}\\

Many thanks to Misha Ryskin and Valery Khoze for a number of  very enlightening discussions and clarifications on this topic and for encouraging me to write  this note. Also thanks for having provided me with  estimates from the KMR model.   In addition, thanks to the members of ATLAS-ALFA group with whom I have discussed this topic many times.  

\printbibliography 

@article{Good:1960ba,
    author = "Good, M. L. and Walker, W. D.",
    title = "{Diffraction disssociation of beam particles}",
    doi = "10.1103/PhysRev.120.1857",
    journal = "Phys. Rev.",
    volume = "120",
    pages = "1857--1860",
    year = "1960"
}

@article{ALICE:2012fjm,
    author = {{ALICE Collaboration}},
    title = "{Measurement of inelastic, single- and double-diffraction cross sections in proton--proton collisions at the LHC with ALICE}",
    eprint = "1208.4968",
    archivePrefix = "arXiv",
    primaryClass = "hep-ex",
    reportNumber = "CERN-PH-EP-2012-238",
    doi = "10.1140/epjc/s10052-013-2456-0",
    journal = "Eur. Phys. J. C",
    volume = "73",
    number = "6",
    pages = "2456",
    year = "2013"
}

@article{TOTEM:2013vij,
    author = {{TOTEM Collaboration}},
    title = "{Luminosity-independent measurements of total, elastic and inelastic cross-sections at $\sqrt{s} = 7$ TeV}",
    reportNumber = "TOTEM-2012-004, CERN-PH-EP-2012-353",
    doi = "10.1209/0295-5075/101/21004",
    journal = "EPL",
    volume = "101",
    number = "2",
    pages = "21004",
    year = "2013"
}

@article{TOTEM:2012oyl,
    author = {{TOTEM Collaboration}},
    title = "{Luminosity-Independent Measurement of the Proton-Proton Total Cross Section at $\sqrt{s}=8$  TeV}",
    reportNumber = "TOTEM-2012-005, CERN-PH-EP-2012-354",
    doi = "10.1103/PhysRevLett.111.012001",
    journal = "Phys. Rev. Lett.",
    volume = "111",
    number = "1",
    pages = "012001",
    year = "2013"
}

@article{TOTEM:2017asr,
    author = {{TOTEM Collaboration}},
    title = "{First measurement of elastic, inelastic and total cross-section at $\sqrt{s}=13$ TeV by TOTEM and overview of cross-section data at LHC energies}",
    eprint = "1712.06153",
    archivePrefix = "arXiv",
    primaryClass = "hep-ex",
    reportNumber = "CERN-EP-2017-321, CERN-EP-2017-321-V2",
    doi = "10.1140/epjc/s10052-019-6567-0",
    journal = "Eur. Phys. J. C",
    volume = "79",
    number = "2",
    pages = "103",
    year = "2019"
}

@article{Ostapchenko:2004ss,
    author = "Ostapchenko, S.",
    editor = "Grieder, P. K. F. and Pattison, B. and Resvanis, L. K.",
    title = "{QGSJET-II: Towards reliable description of very high energy hadronic interactions}",
    eprint = "hep-ph/0412332",
    archivePrefix = "arXiv",
    doi = "10.1016/j.nuclphysbps.2005.07.026",
    journal = "Nucl. Phys. B Proc. Suppl.",
    volume = "151",
    pages = "143--146",
    year = "2006"
}

@article{ATLAS:2014vxr,
    author = {{ATLAS Collaboration}},
    title = "{Measurement of the total cross section from elastic scattering in pp collisions at $\sqrt{s}=7$ TeV with the ATLAS detector}",
    eprint = "1408.5778",
    archivePrefix = "arXiv",
    primaryClass = "hep-ex",
    reportNumber = "CERN-PH-EP-2014-177",
    doi = "10.1016/j.nuclphysb.2014.10.019",
    journal = "Nucl. Phys. B",
    volume = "889",
    pages = "486--548",
    year = "2014"
}

@article{ATLAS:2016ikn,
    author = {{ATLAS Collaboration}},
    title = "{Measurement of the total cross section from elastic scattering in $pp$ collisions at $\sqrt{s}=8$ TeV with the ATLAS detector}",
    eprint = "1607.06605",
    archivePrefix = "arXiv",
    primaryClass = "hep-ex",
    reportNumber = "CERN-EP-2016-158",
    doi = "10.1016/j.physletb.2016.08.020",
    journal = "Phys. Lett. B",
    volume = "761",
    pages = "158--178",
    year = "2016"
}

@article{ATLAS:2022mgx,
    author = {{ATLAS Collaboration}},
    title = "{Measurement of the total cross section and $\rho$-parameter from elastic scattering in $pp$ collisions at $\sqrt{s}=13$ TeV with the ATLAS detector}",
    eprint = "2207.12246",
    archivePrefix = "arXiv",
    primaryClass = "hep-ex",
    reportNumber = "CERN-EP-2022-129",
    month = "7",
    year = "2022"
}

@article{ATLAS:2011zrx,
    author = {{ATLAS Collaboration}},
    title = "{Measurement of the Inelastic Proton-Proton Cross-Section at $\sqrt{s}=7$ TeV with the ATLAS Detector}",
    eprint = "1104.0326",
    archivePrefix = "arXiv",
    primaryClass = "hep-ex",
    reportNumber = "CERN-PH-EP-2011-047",
    doi = "10.1038/ncomms1472",
    journal = "Nature Commun.",
    volume = "2",
    pages = "463",
    year = "2011"
}

@article{ATLAS:2016ygv,
    author = {{ATLAS Collaboration}},
    title = "{Measurement of the Inelastic Proton-Proton Cross Section at $\sqrt{s} = 13$  TeV with the ATLAS Detector at the LHC}",
    eprint = "1606.02625",
    archivePrefix = "arXiv",
    primaryClass = "hep-ex",
    reportNumber = "CERN-EP-2016-140",
    doi = "10.1103/PhysRevLett.117.182002",
    journal = "Phys. Rev. Lett.",
    volume = "117",
    number = "18",
    pages = "182002",
    year = "2016"
}

@article{CMS:2012gek,
    author = {{CMS Collaboration}},
    title = "{Measurement of the Inelastic Proton-Proton Cross Section at $\sqrt{s}=7$ TeV}",
    eprint = "1210.6718",
    archivePrefix = "arXiv",
    primaryClass = "hep-ex",
    reportNumber = "CMS-FWD-11-001, CERN-PH-EP-2012-293",
    doi = "10.1016/j.physletb.2013.03.024",
    journal = "Phys. Lett. B",
    volume = "722",
    pages = "5--27",
    year = "2013"
}

@article{CMS:2018mlc,
    author = {{CMS Collaboration}},
    title = "{Measurement of the inelastic proton-proton cross section at $ \sqrt{s}=13 $ TeV}",
    eprint = "1802.02613",
    archivePrefix = "arXiv",
    primaryClass = "hep-ex",
    reportNumber = "CMS-FSQ-15-005, CERN-EP-2018-004",
    doi = "10.1007/JHEP07(2018)161",
    journal = "JHEP",
    volume = "07",
    pages = "161",
    year = "2018"
}

@article{Kaidalov:1973tc,
    author = "Kaidalov, A. B. and Khoze, Valery A. and Pirogov, Yu. F. and Ter-Isaakyan, N. L.",
    title = "{On determination of the triple pomeron coupling from the ISR data}",
    doi = "10.1016/0370-2693(73)90652-7",
    journal = "Phys. Lett. B",
    volume = "45",
    pages = "493--496",
    year = "1973"
}

@book{Collins:1977jy,
    author = "Collins, P. D. B.",
    title = "{An Introduction to Regge Theory and High-Energy Physics}",
    doi = "10.1017/CBO9780511897603",
    isbn = "978-0-521-11035-8",
    publisher = "Cambridge Univ. Press",
    address = "Cambridge, UK",
    series = "Cambridge Monographs on Mathematical Physics",
    month = "5",
    year = "2009"
}

@article{Khoze:2020plm,
    author = "Khoze, V. A. and Martin, A. D. and Ryskin, M. G.",
    title = "{Dynamics of diffractive dissociation}",
    eprint = "2012.07967",
    archivePrefix = "arXiv",
    primaryClass = "hep-ph",
    reportNumber = "IPPP/20/65",
    doi = "10.1140/epjc/s10052-021-08953-9",
    journal = "Eur. Phys. J. C",
    volume = "81",
    number = "2",
    pages = "175",
    year = "2021"
}

@article{CMS:2015inp,
    author = {{CMS Collaboration}},
    title = "{Measurement of diffraction dissociation cross sections in pp collisions at $\sqrt{s}$ = 7 TeV}",
    eprint = "1503.08689",
    archivePrefix = "arXiv",
    primaryClass = "hep-ex",
    reportNumber = "CMS-FSQ-12-005, CERN-PH-EP-2015-062",
    doi = "10.1103/PhysRevD.92.012003",
    journal = "Phys. Rev. D",
    volume = "92",
    number = "1",
    pages = "012003",
    year = "2015"
}

@article{Khoze:2014aca,
    author = "Khoze, V. A. and Martin, A. D. and Ryskin, M. G.",
    title = "{Elastic scattering and Diffractive dissociation in the light of LHC data}",
    eprint = "1402.2778",
    archivePrefix = "arXiv",
    primaryClass = "hep-ph",
    reportNumber = "IPPP-14-14, DCPT-14-28",
    doi = "10.1142/S0217751X1542004X",
    journal = "Int. J. Mod. Phys. A",
    volume = "30",
    number = "08",
    pages = "1542004",
    year = "2015"
}

@article{ATLAS:2012djz,
    author = {{ATLAS Collaboration}},
    title = "{Rapidity gap cross sections measured with the ATLAS detector in $pp$ collisions at $\sqrt{s}=7$ TeV}",
    eprint = "1201.2808",
    archivePrefix = "arXiv",
    primaryClass = "hep-ex",
    reportNumber = "CERN-PH-EP-2011-220",
    doi = "10.1140/epjc/s10052-012-1926-0",
    journal = "Eur. Phys. J. C",
    volume = "72",
    pages = "1926",
    year = "2012"
}

@article{TOTEM:2017sdy,
    author = {{TOTEM Collaboration}},
    title = "{First determination of the ${\rho }$ parameter at ${\sqrt{s} = 13}$ TeV: probing the existence of a colourless C-odd three-gluon compound state}",
    eprint = "1812.04732",
    archivePrefix = "arXiv",
    primaryClass = "hep-ex",
    reportNumber = "CERN-EP-2017-335, CERN-EP-2017-335-v3",
    doi = "10.1140/epjc/s10052-019-7223-4",
    journal = "Eur. Phys. J. C",
    volume = "79",
    number = "9",
    pages = "785",
    year = "2019"
}

@article{TOTEM:2013lle,
    author = {{TOTEM Collaboration}},
    title = "{Measurement of proton-proton elastic scattering and total cross-section at S**(1/2) = 7-TeV}",
    reportNumber = "TOTEM-2012-002, CERN-PH-EP-2012-239",
    doi = "10.1209/0295-5075/101/21002",
    journal = "EPL",
    volume = "101",
    number = "2",
    pages = "21002",
    year = "2013"
}

@article{TOTEM:2013dtg,
    author = {{TOTEM Collaboration}},
    title = "{Measurement of proton-proton inelastic scattering cross-section at $\sqrt{s}$ = 7 TeV}",
    reportNumber = "TOTEM-2012\textendash{}003, CERN-PH-EP-2012-352",
    doi = "10.1209/0295-5075/101/21003",
    journal = "EPL",
    volume = "101",
    number = "2",
    pages = "21003",
    year = "2013"
}

\end{document}